\pdfoutput=1
\documentclass[%
reprint,
superscriptaddress,
groupedaddress,
nofootinbib,
nobibnotes,
showkeys,
 amsmath,amssymb,
 aps,
 prl,
floatfix,
]{revtex4-1}
\usepackage{amsmath}
\usepackage{graphicx}
\usepackage{aas_macros}
\usepackage[capitalise]{cleveref}

\begin{document}
\newcommand{\rateEarly}{$0.2^{+1.0}_{-0.1}\,\textrm{yr}^{-1}$}
\newcommand{\rateDesign}{$14.2^{+80.5}_{-10.7}\,\textrm{yr}^{-1}$}

\title{Precise LIGO Lensing Rate Predictions for Binary Black Holes}

\bibliographystyle{apsrev4-1}

\author{Ken K.~Y. Ng}%
\email{kwan-yeung.ng@ligo.org}
\affiliation{Department of Physics, The Chinese University of Hong Kong, Shatin N.T., Hong Kong}%
\author{Kaze W.~K. Wong}%
\email{wang-kei.wong@ligo.org}
\affiliation{Department of Physics, The Chinese University of Hong Kong, Shatin N.T., Hong Kong}%
\author{Tjonnie G.~F. Li}%
\affiliation{Department of Physics, The Chinese University of Hong Kong, Shatin N.T., Hong Kong}%
\author{Tom Broadhurst}%
\affiliation{Department of Theoretical Physics, University of Basque Country UPV/EHU, Bilbao, Spain}%
\affiliation{IKERBASQUE, Basque Foundation for Science, Bilbao, Spain}%

\date{\today}

\begin{abstract}

We show how LIGO is expected to detect coalescing binary black holes at $z>1$, that are lensed by the intervening galaxy population.
Gravitational magnification, $\mu$, strengthens gravitational wave signals by $\sqrt{\mu}$, without altering their frequencies, which if unrecognised leads to an underestimate of the event redshift and hence an overestimate of the binary mass.
High magnifications can be reached for coalescing binaries because the region of intense gravitational wave emission during coalescence is so small ($\sim$100km), permitting very close projections between lensing caustics and gravitational-wave events.
Our simulations incorporate accurate waveforms convolved with the LIGO power spectral density.
Importantly, we include the detection dependence on sky position and orbital orientation, which for the LIGO configuration translates into a wide spread in observed redshifts and chirp masses.
Currently we estimate a detectable rate of lensed events \rateEarly{}, that rises to \rateDesign{}, at LIGO's design sensitivity limit, depending on the high redshift rate of black hole coalescence.

\end{abstract}

\maketitle

\section{Introduction}

The brightest and most distant sources in the Universe are often magnified by lensing.
In particular, infrared sky surveys of star forming galaxies have established that lensing by intervening galaxies accounts for the brightest decade in observed mid-IR flux, with magnifications reaching a factor of $50^{+100}_{-4}$, based on modelling each individual case where multiple images are visible \citep{2013ApJ...762...59W, 2013ApJ...779...25B, 2016ApJ...823...17N, 2013ApJ...767..132H, 2016ApJ...826..112S, 2016ApJ...824...86S}.
The maximal magnifications of such star forming galaxies is limited by the size of the bright star forming regions because the smaller the source, the closer it can lie in projection near the lensing caustic.
The larger the source the lower the overall magnification, saturating at a maximum magnification for sources that straddle a caustic.
An extreme example is the well-known IRAS F10214 ($z=2.2$) that is magnified $\sim 100$ times \citep{1995ApJ...450L..41B} where a compact central IR emission is projected on the maximally magnifying cusp caustic of an intervening elliptical lensing galaxy.

This size-dependent magnification effect can be more extreme for gravitational-waves (GW) sources that are detectable by ground-based interferometers, such as LIGO \citep{2015CQGra..32g4001L}, because of the tiny region of intense wave emission.
Considering a typical galaxy lens for which the size of the Einstein ring is $\sim 1\mathrm{kpc}$, gravitational-wave sources such as mergers of compact objects, which are $\sim 100 \mathrm{km}$ in size, can be treated the geometrical-optics limit.
With the recent gravitational-wave detections \citep{2016PhRvL.116f1102A,2016PhRvL.116x1103A} and the subsequent prediction of many more similar events \citep{2016PhRvX...6d1015A}, it is natural to wonder how many of these detections will be lensed.

The effect of gravitational magnification, $\mu$, on GWs emitted by a compact source is to enhance the detected strain amplitude by $\sqrt{\mu}$ without changing the observed frequency structure of the waveform \citep{1996PhRvL..77.2875W, 2017PhRvD..95d4011D} since lensing in the geometrical-optics limit is ``achromatic''.
This means the distance inferred to a lensed event is degenerate with the unknown magnification, unless lensing can be excluded.
Therefore, a distant magnified source can be equally inferred to be luminous and relatively nearby if the role of lensing is unknown.
Furthermore, GW events from binary black hole coalescence are not accurately localised on the sky in the absence of an associated electromagnetic signature, and hence it is not possible to exclude lensing of such events, taken individually.
Finally, the degeneracy between the chirp mass and the redshift implies uncertainty regarding the intrinsic chirp mass of an unidentified GW event.
Cosmological stretching of the waveform with source redshift, $z_s$, can be simply compensated by increased binary orbital frequency, corresponding to a lower intrinsic chirp mass, so that the observed ``chirp'' mass $\mathcal{M}_c$, is larger than its intrinsic value: $\mathcal{M}_c=(1+z_s)\mathcal{M}_0$.
This redshift-chirp mass degeneracy can only be broken if the redshift can be measured through some auxiliary measurement and lensing is excluded.

These degeneracies can be understood by considering the signal-to-noise ratio $\rho$ which scales to leading order as 
\begin{align}
	\rho' \sim \sqrt{\mu} \Theta \mathcal{M}_c^{5/6}/d_L.
\end{align}
The ``geometrical'' term $\Theta$ depends strongly on the poorly measured sky position and orbital orientation through 
\begin{align}
\Theta \equiv 2\left[ F_{{}+{}}^2\left(1+\cos^2i\right)^2 + 4F_{\times}^2\cos^2i\right]^{1/2},
\end{align}
where the antenna pattern functions $F_+$ and $F_\times$ are given by
\begin{align}
F_{{}+{}} \equiv {1\over2}\left(1+\cos^2\theta\right)\cos2\phi\cos2\psi - \cos\theta\sin2\phi\sin2\psi,\\
F_\times \equiv {1\over2}\left(1+\cos^2\theta\right)\cos2\phi\sin2\psi + \cos\theta\sin2\phi\cos2\psi,
\end{align}
where $\theta, \phi, \iota$ and $\psi$ are the sky location angles, orbital orientation and polarization angle respectively.
This geometrical term ranges over $0<\Theta< 4$, and peaks for sources directly overhead LIGO \citep{1996PhRvD..53.2878F}.
However, $\Theta$ is subject to a considerable uncertainty due to the poor angular resolution of interferometric detectors \citep{2016PhRvX...6d1015A}.

Calculations of gravitational wave lensing first focussed on neutron star mergers \citep{1996PhRvL..77.2875W} including future space missions \citep{2011MNRAS.415.2773S, 2014JCAP...10..080B} and the related precision on $H_0$ \citep{1986Natur.323..310S, 2013arXiv1307.2638N, 2012PhRvD..86d3011D} and more recently extended to binary black holes (BBH) \citep{2017PhRvD..95d4011D}, given the LIGO detections \citep{2016PhRvX...6d1015A}.
Here we enhance the precision of strong lensing predictions through Monte Carlo simulation, and introduce the significant role of the angular dependence, $\Theta$, as this plays a large role which cannot be ignored in the case of LIGO events, as we demonstrate below.
We also take care to make detailed simulations of coalescing waveform in frequency space convolved with the frequency bandpass for LIGO so that the GW signal is accurately predicted.
We rely on the signal-to-noise ratio as our signal discriminator to which we add lensing that is known to be dominated by massive early type galaxies, confirming the lensing ``optical depth'' predictions of \citep{2000ApJ...536..571B}.

\section{Generating Binary Black Hole Events} \label{SimSignal}
As the successful GW observations are all BBH events, we focus here on BBH only.
A gravitational waveform depends on the component masses $m_1$ and $m_2$, spins $\vec{S}_1$ and $\vec{S}_2$, the redshift $z$ and angular detector response $\Theta$
In this work, we assume that the black hole spins are aligned or anti-aligned with the orbital angular moment so that spin effects can be characterised by a single effective spin $\chi_{\rm eff} = (m_1 S_{1z} + m_2 S_{2z})/(m_1 + m_2)$.

We take the mass of BHs to be distributed according to the relatively detailed population synthesis simulations of \textcite{2013ApJ...779...72D}, with effective spin distributed uniformly within $\chi_{\rm eff} \in \left( -1,1 \right)$.
Delay time to coalescence and evolution of BBH are highly uncertain.
For simplicity, we adopt a redshift distribution, which is also the merger rate density, of the form $P(z) \sim (1+z)^{\alpha}$ with $z_{\mathrm{max}}<2.5$, where $\alpha\simeq 3$ is approximately the measured evolution of the integrated star formation rate at low-z \citep{1990MNRAS.242..318S} and for which the formation of massive BH stellar progenitors may be expected to follow approximately, depending on the unknown details of BBH binary formation and coalescence delay time.
To convert redshifts to distances we adopt Planck weighted cosmological parameters: $H_0=67.8\,\text{km/s/Mpc}$, $\Omega_{\mathrm{m}}=0.306$, $\Omega_{\Lambda}=0.694$ and $\Omega_k=0$ \citep{2016A&A...594A..13P}.
We further assume that the sky location and orbital orientation are uniformly distributed on the corresponding unit spheres.

From the source distributions we calculate the signal-to-noise ratio (SNR) $\rho$ using the noise-weighted inner product of the waveform
\begin{equation}
	\rho^2 = \int^{f_{\mathrm{max}}}_{f{\mathrm{min}}} \frac{h(f)h^*(f)}{S_n(f)} df,
\end{equation}
where $h(f)$ is the strain signal in frequency domain and $S_n(f)$ is the noise power spectrum.
We use the inspiral-merger-ringdown phenomenological waveform model from \cite{2016PhRvD..93d4007K,2016PhRvD..93d4006H} to simulate the gravitational-wave strain from a BBH merger.
Moreover, we assume the publicly available O1 noise power spectral density \citep{2015JPhCS.610a2021V} as a proxy for the sensitivity of the LIGO detectors in the near future.
Finally, a signal is classified as ``detectable'' when the SNR is above the SNR threshold $\rho \geq \rho_{\rm th} = 8$.

\section{Lensing Optical Depth} \label{Lens}
An ``optical depth'' for lensing $\tau (z)$ can be defined as the fraction of the sky that is enhanced in area by the lens magnification \cite{1984ApJ...284....1T}, so that for a source at redshift $z_s$, the probability of being lensed by magnification $\mu>\mu_{\mathrm{min}}$ can be defined as 
\begin{equation}
	P(\mu>\mu_{\mathrm{min}}, z_s) = \tau(z_s) \cdot P(\mu>\mu_{\mathrm{min}} \vert z_s).
\end{equation}
For early-type galaxies the optical depth has a simple form by relating the internal galaxy velocity dispersion to the isothermal mass density profile \cite{1991MNRAS.253...99F}:
\begin{equation}
	\tau(z_s) = \frac{F_*}{30}\left[ \frac{d_C(z_s)}{cH_0^{-1}} \right]^3,
\end{equation}
where $d_C$ is the transverse comoving distance, $H_0$ is the Hubble's constant and the normalisation $F_*$ of the optical depth has been determined to be $\simeq 0.05-0.07$\citep{1991MNRAS.253...99F, 1996PhRvL..77.2875W}.
The probability distribution of magnifications is given by the universal form $P(\mu>2) \propto \mu^{-3}$ in the strong lensing regime and this is usually integrated above a lower limit, $\mu_{\mathrm{min}}=2$ to encompass all multiply lensed images for the isothermal mass distribution for which the outer image has a lower limiting magnification, $\mu=2$.
We are insensitive to this limit, as we show below, because our predicted rates for detectable events relate to much higher magnifications.

\section{Calculation of expected lensing rates}\label{result}
We simulate $N$ events using the distributions of BBH masses, redshifts and geometrical factor calculated as described above.
From the simulated events, we obtain a distribution of intrinsic SNR $\rho$.
We now calculate the intrinsic differential rate at each $d\mathcal{M} dz$
\begin{equation}
	\frac{d^2 R}{d\mathcal{M} dz} = A \frac{d^2 P}{d\mathcal{M} dz},
\end{equation}
where $d^2 P/d\mathcal{M} dz$ is the probability density of intrinsic events.
The normalization $A$ depends on the comoving volume $V$ and redshift distribution through
\begin{equation}
	A = \int_{0}^{z_{\mathrm{max}}} P(z) \mathcal{R}(0) \frac{dV(z)}{dz} dz,
\end{equation}
where we take the local merger rate density $\mathcal{R}(0)$ to be $36\mathrm{ Gpc}^{-3} \mathrm{ yr^{-1}}$ from Ref.~\citep{2015ApJ...806..263D}.
We approximate the error of local merger rate density by the 90\% confidence interval of inferred rate in LIGO's O1 run \citep{2016PhRvX...6d1015A}.

\begin{figure}[h!]
	\centering
	\includegraphics[width=\columnwidth]{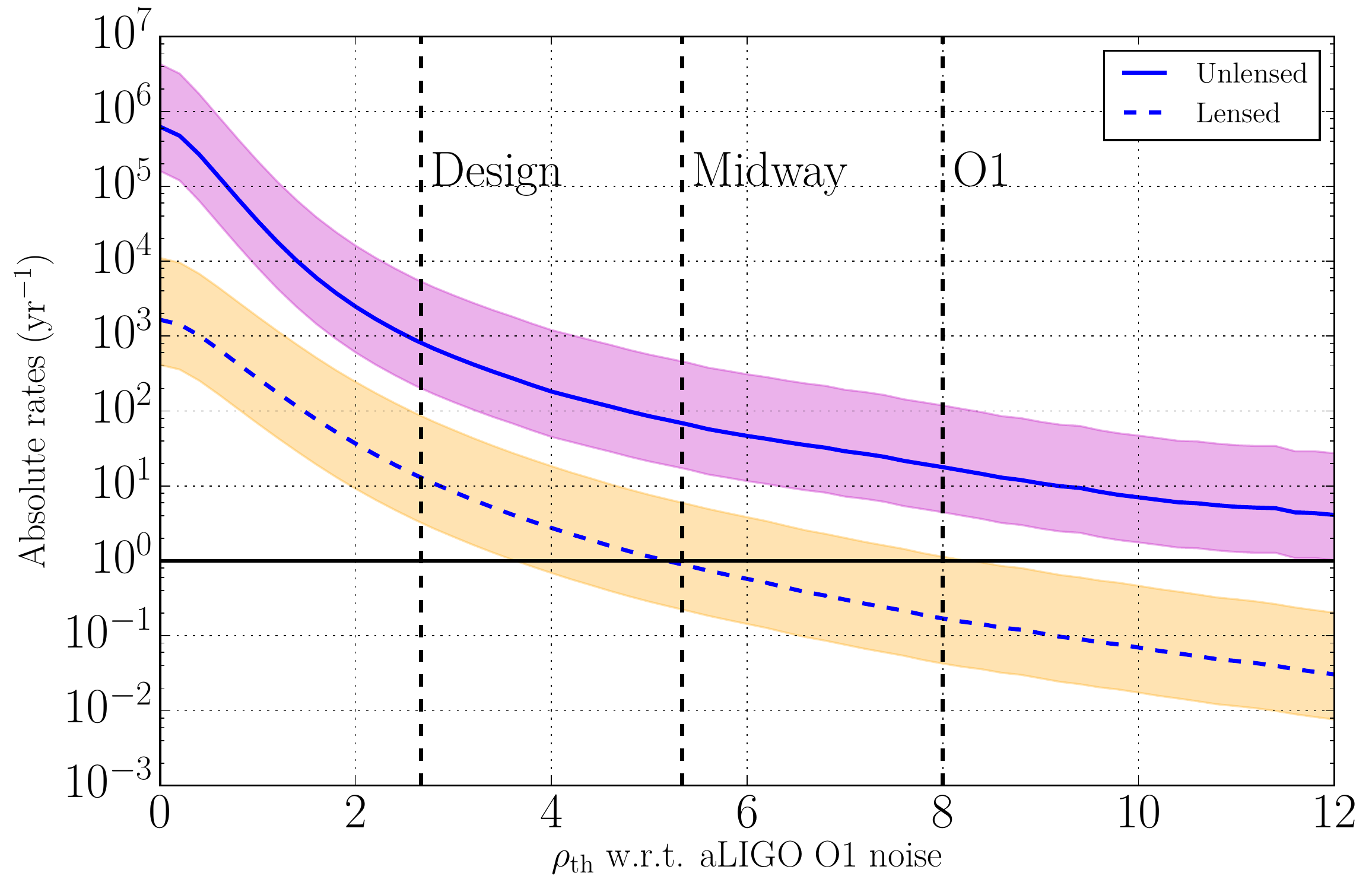}
	\caption{
		Predicted rates in different stages of LIGO using the mass function from \textcite{2013ApJ...779...72D}.
		The solid line shows the overall rate of events and the dashed line shows the rate of lensed events.
		Even though the rate of lensed events only contributes a small fraction of the overall rate of events, lensed events may be observed frequently in the Advanced detector era.
	}
	\label{rate}
\end{figure}

To determine the rate of lensed signals over some SNR threshold $\rho_{\mathrm{th}}$, the differential lensed rate is equivalent to the ratio of number density between lensed events $d^2 N_{L}(>\rho_{\mathrm{th}})/d\mathcal{M} dz$ and intrinsic events $d^2 N/d\mathcal{M} dz$, multiplied by the absolute differential rate at each $d\mathcal{M} dz$ bin.
Weighted by optical depth $\tau(z)$, we can estimate the number density of lensed events in our simulation.
The total lensed rate is then given by
\begin{widetext}
	\begin{align}
		R_{L}(>\rho_{\mathrm{th}}) = \int_{\mathcal{M}_{\mathrm{min}}}^{\mathcal{M}_{\mathrm{max}}} \int_{0}^{z_{\mathrm{max}}} \frac{d^2 R}{d\mathcal{M} dz} \frac{d^2 N_{L}(>\rho_{\mathrm{th}})/d\mathcal{M} dz}{d^2 N/d\mathcal{M} dz} dz d\mathcal{M}.
	\end{align}
\end{widetext}
Fig. \ref{rate} shows how the lensing rate varies increases with lower limiting $\rho_{\mathrm{th}}$, where we indicate the current and expected future sensitivities of LIGO by the vertical dashed lines.
The rate is found to be \rateEarly{} at LIGO's current sensitivity, and rises to \rateDesign{} at the design sensitivity.

\begin{figure}[h!]
	\centering
	\includegraphics[width=\columnwidth]{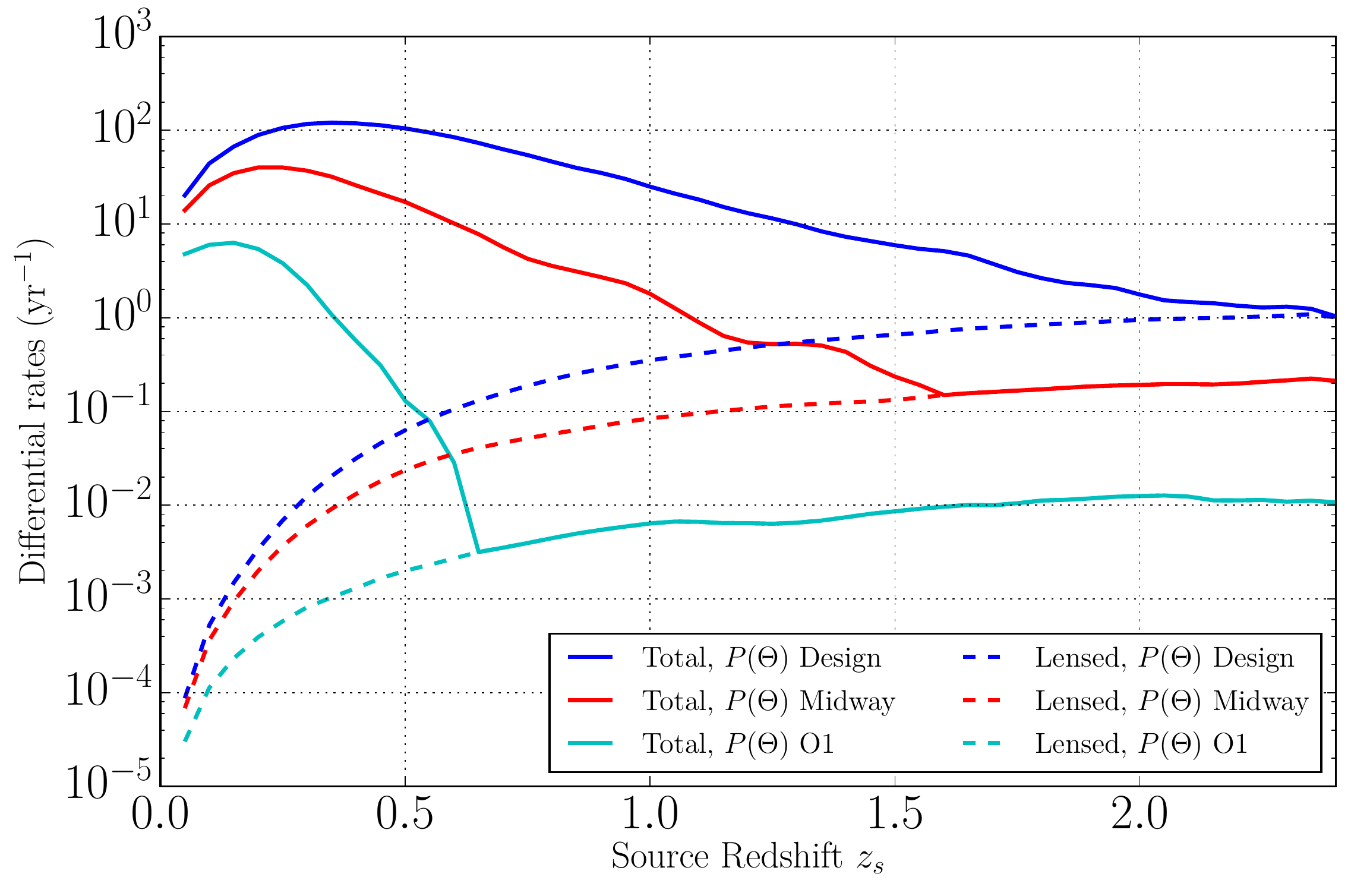}
	\caption{
		The differential rate as a function of the source redshift $z_s$.
		The solid lines show the differential rate of overall events and the dashed lines show the differential rate of lensed events.
		The colors represent different stages of LIGO.
	}
	\label{rate_z}
\end{figure}

Fig. \ref{rate_z} shows the differential rate as function of the source redshift at different stages of LIGO.
Lensing starts to dominate the rate for $z>0.6$ at current sensitivity and rises to $z>2$ for the design sensitive of LIGO, and causes a sharp transition in the differential rate.
However, the large uncertainty of the distance measurements means that it is unlikely to observe this transition through gravitational-wave observation alone.

\begin{figure}[h!]
	\centering
	\includegraphics[width=\columnwidth]{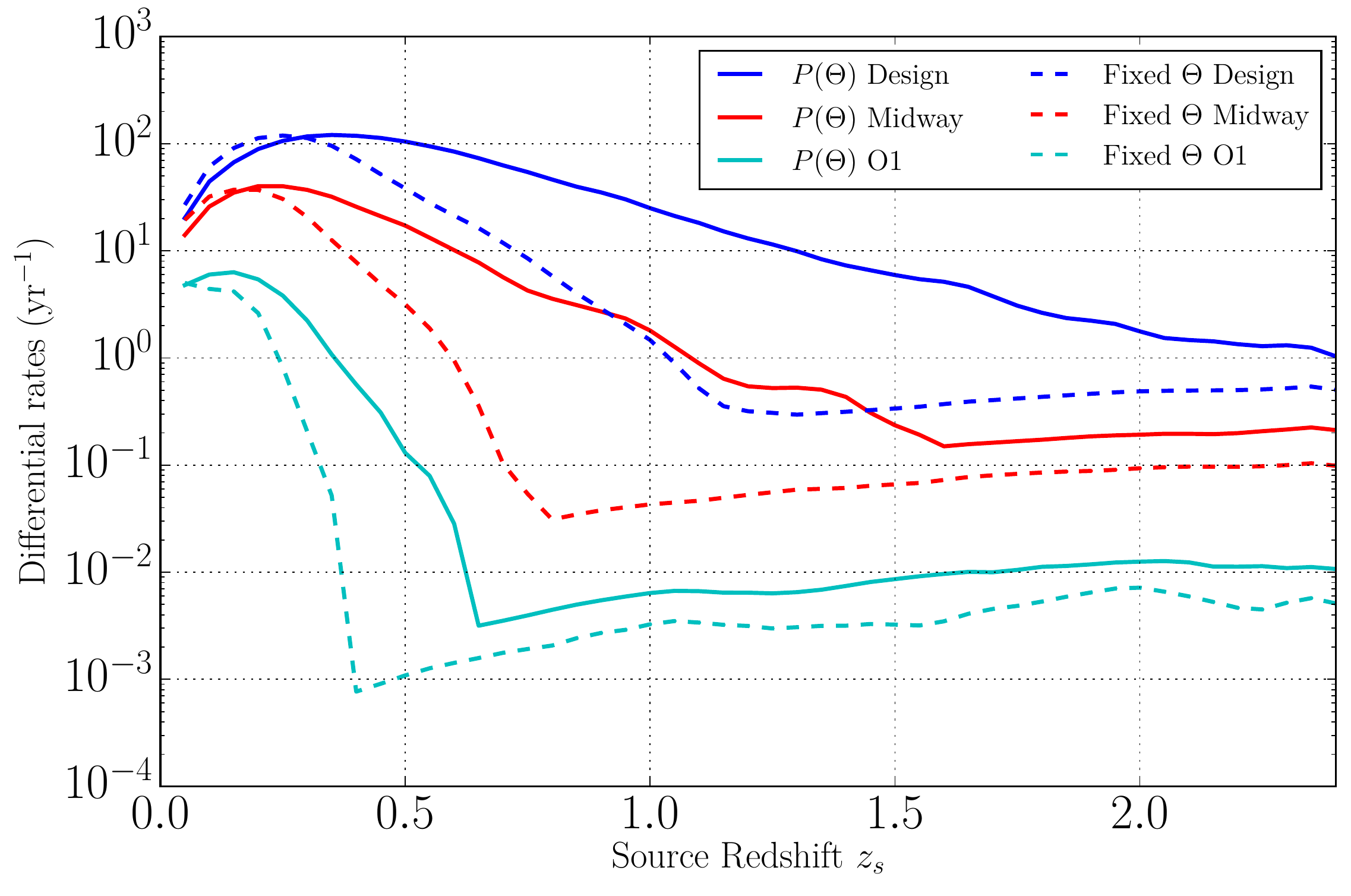}
	\caption{
		The differential rate as a function of the source redshift $z_s$ for a realistic (solid) and a fixed (dashed) $\Theta$ distributions at different stages of the LIGO's sensitivity.
	}
	\label{fig:rate_z_theta}
\end{figure}

Previous work in Ref.~\citep{2017PhRvD..95d4011D} adopts the simple sky-averaged value of $\Theta_{\rm av}=1.6$ for LIGO, which overlooks the wide range of detection depth on the sky, which is maximal for sources located overhead a plane defined by the LIGO detectors at any given time.
Fig. \ref{fig:rate_z_theta} shows the differential rate as a function of redshift $z$ for a fixed $\Theta_{\rm av}=1.6$ and for a realistic $\Theta$.
When considering the full effect of the geometrical factor $\Theta$, the population of detectable sources increases significantly for redshifts higher than $z_s > 0.2$, and thereby increasing the overall rate of detectable sources.
In particular, the less massive sources at higher redshift lie in the region with higher detector response, so that a larger $\Theta$ can compensate for intrinsically low amplitude.

\begin{figure}[h!]
	\centering
	\includegraphics[width=\columnwidth]{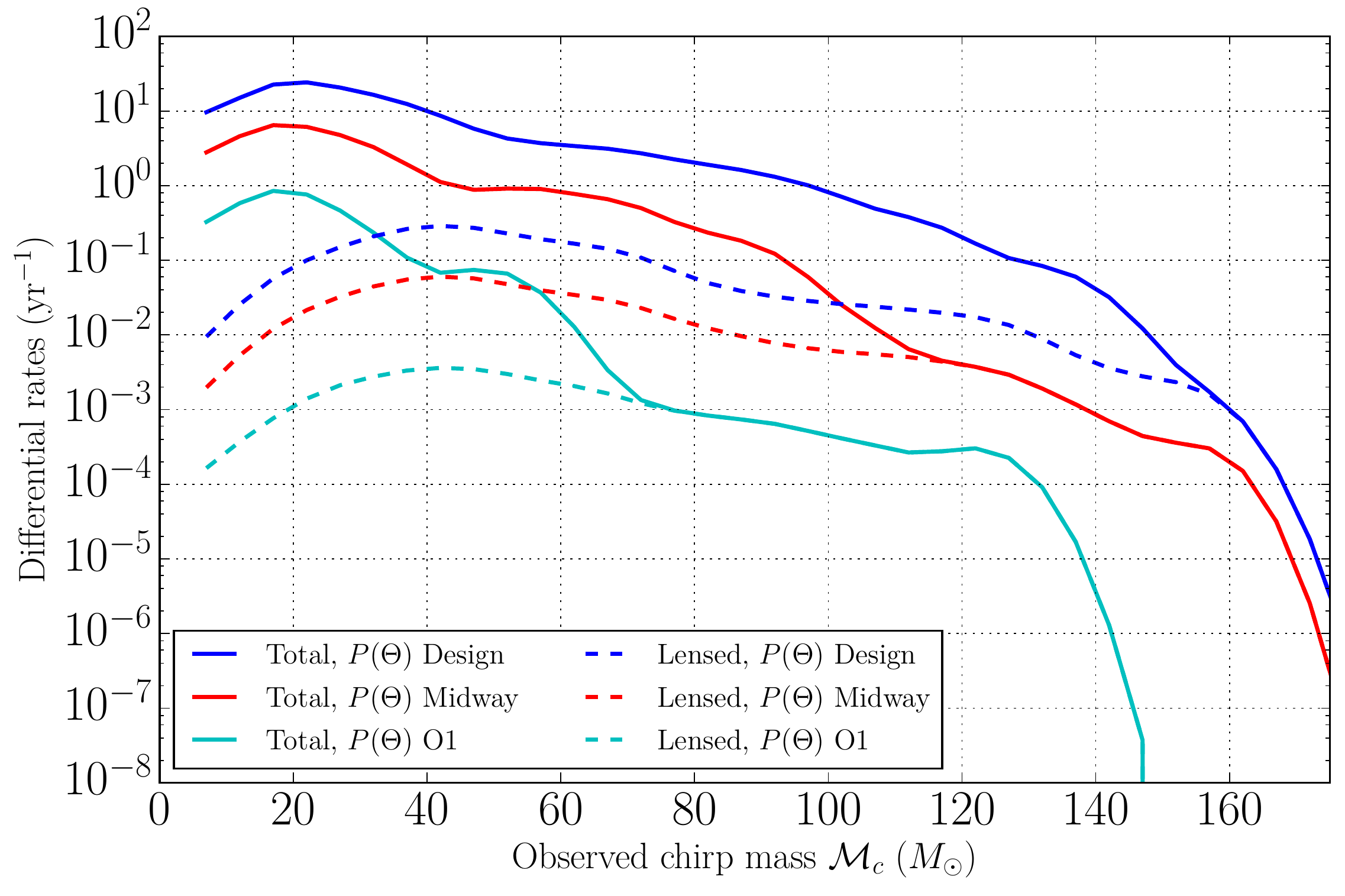}
	\caption{
		The differential rate as a function of the observed chirp mass $\mathcal{M}_c$ for all events (solid) and for lensed events (dashed) at different stages of LIGO's sensitivity.
		Lensing introduces a sharp transition in the differential rate that corresponds to the cut-off due to the intrinsic mass distribution.
	}
	\label{fig:rate_mchirp}
\end{figure}

Finally, Fig. \ref{fig:rate_mchirp} shows the cumulative rate as a function of the observed chirp mass $\mathcal{M}_c$.
Without lensing the observed chirp mass distribution at the current sensitivity extends to over $\mathcal{M}_c > 60\,M_{\odot}$ due to the cosmological redshift.
Once lensing is considered, the observed chirp mass at the current sensitivity can extend all the way to $\mathcal{M}_c > 140\,M_\odot$.
Similar behaviour, but at higher values for the chirp mass, can be seen as LIGO's sensitivity improves.
Lensing introduces a sharp transition in the differential rate that corresponds to the cut-off due to the intrinsic mass distribution.
This sharp transition can be the tell-tale sign of lensing, and reveal information about the intrinsic mass distribution for binary black holes.

\section{Discussion}\label{Discussion}
We have estimated the rate of gravitational-wave signals that are lensed by elliptical galaxies through Monte Carlo simulations of the LIGO detectors.
In particular, we simulated binary black hole events convolved with realistic lensing models in order to estimate fraction of LIGO events that are lensed.
Our results show that by adopting reasonable assumptions about the source population, the observable rate of lensed gravitational-wave signals from binary black holes that we predict for the LIGO detectors is \rateEarly{}, rising to \rateDesign{} at the design sensitivity of LIGO.
Therefore, it is likely to see at least 1 lensed event/yr in the Advanced detector era.

In particular, the low optical depth for lensing is countered by a relatively large magnification bias, as weaker events are strengthened by lens magnification encompassing an enlarged cosmological volume, such that the strongest detections by LIGO in the future may preferentially include lensed events.
Our estimates adopt conservative assumptions regarding the BBH mass function and its evolution so that in principle our rates may be underestimated if the formation rate of massive is much higher at $z<1$ where lensing dominates.
Moreover, the relative rate of lensed to unlensed events can be much higher if the BBH mass function above $\gtrsim 10M\odot$ has a steep slope so that relatively low magnification events can dominate the strongly observed chirp mass events.
We have not investigated micro-lensing, which is expected to be significant near the Einstein radius of massive galaxy lenses where stars and possible substructure in the dark matter may significantly enhance the lensing optical depth.
In this regime it will also be important to examine diffraction effects that will limit the maximal magnification particularly for gravitational waves because of their relatively low frequency \cite{2003ApJ...595.1039T}.

The non-negligible lensed rate opens up the possibility of seeing multiple signal from the same event.
Typically, the first multiply lensed event is the most magnified with the later events being significantly weaker in strength, especially for galaxy lenses that do not typically have a large radial critical curve.
The search for such sub-threshold events of common origin must have the same chirp mass and can rely on prior knowledge of the expected distribution of galaxy scale time delays and relative magnification between lensed signals.
A detailed calculation of these quantities and studies on the ability to statistically identify lensed signals are ongoing and will appear in upcoming work.
Signatures of lensing may also emerge from the population properties of detected events, including high chirp masses with optimal sky plane inclinations, distributed in amplitude along the relatively shallow universal caustic cusp relation for strongly magnified events.

\begin{acknowledgments}
This research has made use of data, software and/or web tools obtained from the LIGO Open Science Center (https://losc.ligo.org), a service of LIGO Laboratory and the LIGO Scientific Collaboration.
LIGO is funded by the U.S. National Science Foundation.
The authors would like to acknowledge the LIGO Data Grid clusters.
T.~J.~Broadhurst gratefully acknowledges the hospitality of the Chinese University of Hong Kong and the Visiting Research Professor Scheme at the University of Hong Kong.
\end{acknowledgments}

\end{document}